\documentclass[aps,prc,showpacs,superscriptaddress,amssymb,amsmath,amsfonts,floatfix,nofootinbib]{revtex4-1}
\usepackage{graphicx}% Include figure files
\usepackage{natbib}
\usepackage[usenames]{color}
\usepackage{epsfig}
\usepackage{amssymb,amsmath}
\usepackage{epstopdf}
\usepackage{subfig}
\usepackage{lineno,hyperref}
\usepackage{enumitem}

\definecolor{grey}{rgb}{0.9,0.9,0.9}
\definecolor{black}{rgb}{0,0,0}

\hypersetup{pdfpagemode=FullScreene,pdfstartview={XYZ null null 1.10}}

\newcommand{\di}{i}

\newcommand{\be}{\begin{eqnarray}}
\newcommand{\ee}{\end{eqnarray}}
\newcommand{\bc}{\begin{center}}
\newcommand{\ec}{\end{center}}
\newcommand{\beq}{\begin{eqnarray}}
\newcommand{\eea}{\end{eqnarray}}

\begin{document}

%%%%%%%%%%%%%%%%%%%%%%%%%%%%%%%%%%%%%%%%%%%%%%%%%%
\title{The role of angle dependent phase rotations of reaction amplitudes  in \mbox{$\eta$ photoproduction on protons  }
\vspace*{0.2cm}}
\newcommand{\Zagreb}[1]
{\affiliation{ Rudjer Bo\v{s}kovi\'{c} Institute, Bijeni\v{c}ka cesta 54, P.O. Box 180, 10002 Zagreb, Croatia}}

\newcommand{\Tuzla}[1]
{\affiliation{University of Tuzla, Faculty of Natural Sciences and Mathematics, Univerzitetska 4, \\ 75000 Tuzla, Bosnia and Herzegovina}}

\newcommand{\Mainz}[1]
{\affiliation{Institut f\"ur Kernphysik, Johannes Gutenberg-Universit\"at Mainz, \\ D-55099 Mainz,Germany}}

\author{A.~\v{S}varc}\thanks{Corresponding author: alfred.svarc@irb.hr} \Zagreb \\
\author{H.~Osmanovi\'{c}} \Tuzla \\
\author{ M.~Had\v{z}imehmedovi\'{c}} \Tuzla \\
\author{ R.~Omerovi\'{c}} \Tuzla \\
\author{ J.~Stahov} \Tuzla \\
\author{ V.~Kashevarov} \Mainz \\
\author{ K.~Nikonov} \Mainz \\
\author{ M.~Ostrick} \Mainz \\
\author{ L.~Tiator \vspace*{0.5cm}} \Mainz \\

%%%%%%%%%%%%%%%%%%%%%%%%%%%%%%%%%%%%%%%%%%%%%%%%%%%%%%%%%%%%%%%%%%

\date{\today }

%%%%%%%%%%%%%%%%%%%%%%%%%%%%%%%%%%%%%%%%%%%%%%%%%%%%%%%%%%
\begin{abstract}
\vspace*{0.5cm}
It has recently been proven that the invariance of observables with respect to angle dependent phase rotations of reaction amplitudes  mixes multipoles changing also their relative strength \cite{Svarc2018}. All contemporary partial wave analyses (PWA) in $\eta$ photoproduction on protons, either energy dependent (ED) \cite{BoGa,Kent,Juelich,MAID} or single energy (SE) \cite{fixed-t}
do not take this effect into consideration.  It is commonly accepted that there exist quite some similarity in the $E0+$ multipole for all PWA, but notable differences in this, but also in remaining partial waves still remain. In this paper we demonstrate that once this phase rotations are properly taken into account, all contemporary ED and SE partial wave analysis become almost identical for the dominant $E0+$  multipole, and the agreement among all other multipoles becomes much better. We also show that the the measured observables are almost equally well reproduced for all PWA, and  the remaining differences among multipoles can be attributed solely to the difference of predictions for unmeasured observables. So, new measurements are needed.

\end{abstract}
%%%%%%%%%%%%%%%%%%%%%%%%%%%%%%%%%%%%%%%%%%%%%%%%%%%%%%%%%%

\maketitle
%%%%%%%%%%%%%%%%%%%%%%%%%%%%%%%%%%%%%%%%%%%%%%%%%%%%%%%%%%%%%%%%%%%%%%%%%%%%%%%%

\begin{figure}[h!] \hspace*{-0.5cm}
\begin{center}
\includegraphics[width=0.85\textwidth]{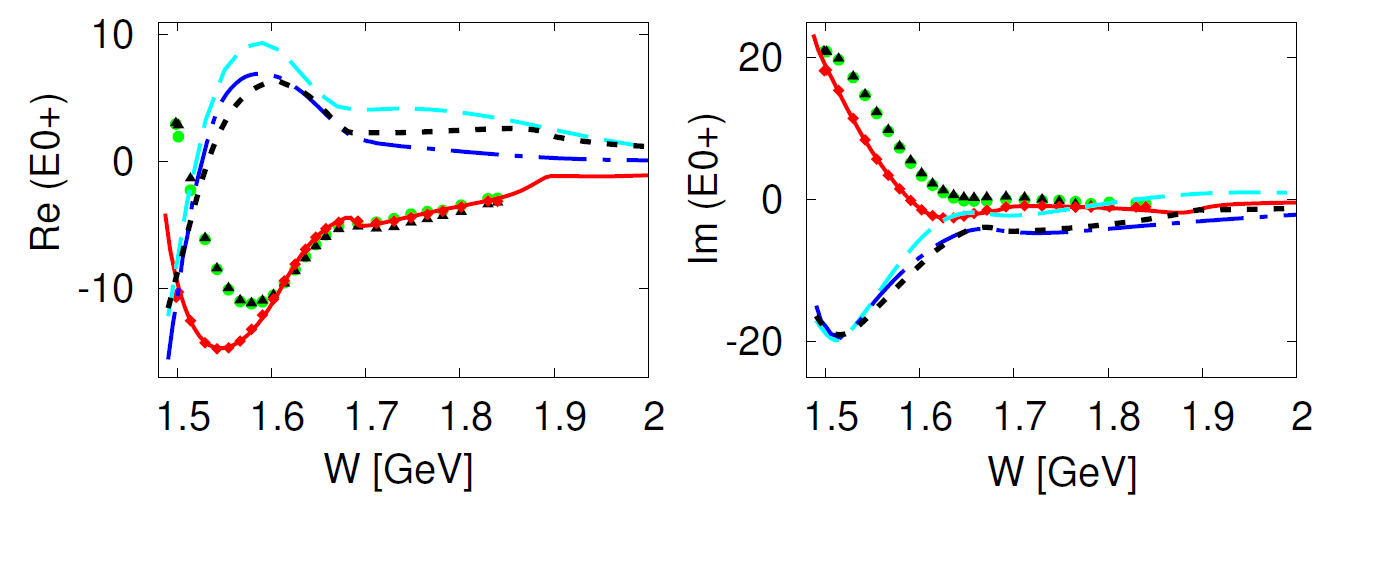}
\caption{\label{E0+non-rotated}(Color online) Comparison of $E0+$  $\eta$ photoproduction multipole for the  Mainz EtaMAID model \cite{MAID} (red, full line),  Bonn-Gatchina calculation \cite{BGmultipoles} (black, short-dashed line), Kent State University calculation \cite{Manleymultipoles} (cyan, long-dashed line),  J\"{u}lich-Bonn model \cite{Doeringmultipoles} (blue, dash-dotted line), and three solutions from SE fixed-t analysis \cite{fixed-t} (discrete symbols).     }
\end{center}
\end{figure}

In recent years, a wealth of new high-precision experimental data dominantly on photo- and electro-production has been measured at various facilities including JLab, MAMI, LEPS, SLAC, and GRAAL for a number of observables with the goal of better understanding the spectrum of $N^*$ and $\Delta$ resonances.
A chain of coupled-channel models including photo- and electro-production have therefore been  developed \cite{BoGa,Kent,Juelich,MAID} with the aim of including the plethora of new data into one, unified overall scheme. Number of channels varied from more than seven in most models to only two  ($\eta$-N and $\eta'$-N)  in ref. \cite{MAID}. A single energy (SE), single-channel method for $\eta$ photo-production based on achieving the continuity of solution by imposing fixed-t analyticity has been recently also added to these theoretical efforts \cite{fixed-t}. As a result,  a number of equivalent sets of biggest partial waves for $\eta$ photo-production  was generated. Now, after decades of research, it is commonly accepted that there exist quite some similarity in the dominant partial wave $E0+$ among all models, but notable differences in this and all other partial waves still remain. The differences were mostly attributed to the difference in model assumptions (number of resonances, dynamics, background treatment, etc.), and in data bases used to constrain the free model-parameters (data selection, weighting, interpolations, data binning, etc.),  and no one suspected that there might be another, fundamental reason why all these calculations disagree at least for the dominant multipole. Hereafter we show that such a reason exists, and that it lies in the inadequate treatment of continuum ambiguity effects which manifest themselves as  angle dependent phase rotations of reaction amplitudes \cite{Svarc2018}.
\\ \\ \indent
As all observables in meson production processes are given in term of bilinears of one amplitude with the complex conjugate of another one, so they are invariant with respect to the energy and angle dependent phase rotation. This invariance is called continuum ambiguity \cite{Atk73,Bow75,Atk85}.  We formalize it in the following  way:
The observables in single-channel reactions are given as a sum of products involving one amplitude
(helicity, transversity, ... ) with the complex conjugate of another one, so that the general form of
any observable is given as ${\cal O}=f˙(H_k \cdot H_l^*)$, where $f$ is a known, well-defined real function.
The direct consequence is that any observable is invariant with respect to the following simultaneous
phase transformation of all \vspace*{0.3cm} amplitudes:
\be
H_k (W,\theta) \rightarrow  \tilde{H_k} (W,\theta)& = & e^{\, i \, \,  \phi(W,\theta) } \cdot H_k (W,\theta) \label{eq:ContAmbGeneralTrafo} \\
 {\rm for \, \, all} \, \, k & = & 1,\cdots,n \nonumber
\ee
\noindent
where n is the number of spin degrees of freedom (n=1 for the 1-dim model, n=2 for pi-N scattering
and n=4 for pseudoscalar meson photo-production), and $\phi(W,\theta)$ is an arbitrary, real function which is the same for all contributing amplitudes.
Without any further physics constraints like unitarity, this real function $\phi(W,\theta)$ is free, and there exist an infinite number of equivalent solutions which give exactly the same set of observables. The invariance with respect to energy dependent phase rotation has been investigated a lot, and  synchronizing phases were introduced and analyzed on the level of partial waves \cite{bnga,Tiator:2011pw,Sandorfi2011,Manley2012}. These rotations can be handled without any problems.  However, almost no attention has been paid to the situation when the arbitrary phase function is angle dependent.  This possibility was mentioned in \cite{Oma81,Gibs2008,Dean,Keaton} where the effect was established, but the discussion was not followed through. The whole deduction chain for understanding the full role of angle dependent phase rotations in continuum ambiguity was finally presented in \cite{Svarc2018}.
\\ \\ \indent
Starting with  Eq.~(\ref{eq:ContAmbGeneralTrafo}),  we focus on  resonance properties of amplitudes $ H_k (W,\theta) $. As resonances are identified with poles of the partial-wave (or multipole) amplitudes,  we must analyze the influence of the continuum ambiguity not
upon helicity or transversity amplitudes, but upon their partial-wave decompositions.
To streamline the study we introduce partial waves in a version simplified with respect to the form found in, for instance,
ref. \cite{Tiator:2011pw}:
\be \label{Eq:PW}
A (W,\theta) & = & \sum^{\infty}_{\ell=0} (2 \ell + 1) A_\ell(W) P_{\ell}(\cos\theta)
\ee
\noindent
where $A(W,\theta)$ is a generic notation for any amplitude $H_k(W,\theta)$, $k=1, \cdots n$.
The complete set of observables remains unchanged when we make the following transformation:
\be \label{Eq:PWrot}
A (W,\theta) \rightarrow \tilde{A} (W,\theta)& = & e^{\, \di \, \,  \phi(W,\theta) }
 \hspace*{5pt} \times \hspace*{5pt}   \sum^{\infty}_{\ell=0} (2 \ell + 1) A_\ell(W) P_{\ell}(\cos\theta) \nonumber \\
\tilde{A} (W,\theta) & = &  \sum^{\infty}_{\ell=0} (2 \ell + 1) \tilde{A}_\ell(W)P_{\ell}(\cos\theta)
\ee
\noindent
We are interested in rotated partial wave amplitudes $\tilde{A}_\ell(W)$, defined by Eq.(\ref{Eq:PWrot}),
and are free to introduce the Legendre decomposition of an exponential function as:
\be \label{Eq:Phaseexpansion}
e^{\, \di \, \,  \phi(W,\theta) } &=& \sum^{\infty}_{\ell=0} L_\ell(W)  P_{\ell}(\cos\theta).
\ee
After some manipulation of the product
$P_\ell(x) P_k(x)$
(see refs.~\cite{Dougall1952,Wunderlich2017}~for details of the summation rearrangement) we obtain:
\be \label{Eq:mixing}
\tilde{A}_\ell(W) &=& \sum_{\ell'=0}^{\infty} L_{\ell'}(W) \, \, \, \cdot \sum_{m=|\ell'-\ell|}^{\ell'+\ell}\langle \ell',0;\ell,0|m,0 \rangle ^2\, \, A_{m}(W)
\ee
where $\langle \ell',0;\ell,0|m,0 \rangle $ is a standard Clebsch-Gordan coefficient.  A similar relation was also derived in ref.~\cite{Gibs2008}.
\\ \\ \indent
To get a better insight into the mechanism of multipole mixing, let us expand Eq.~(\ref{Eq:mixing}) in terms of phase-rotation Legendre coefficients $L_{\ell'}(W)$, and demonstrate that angle dependent phase invariance mixes multipoles: \\
%\onecolumngrid
\begin{align} \label{Eq:mixing-expanded}
 \tilde{A}_{0} (W) &= {L_{0} (W) A_{0} (W) } + L_{1} (W) A_{1} (W) + L_{2} (W) A_{2} (W) + \ldots  \mathrm{,} \\
 \tilde{A}_{1} (W) &= {L_{0} (W) A_{1} (W) } + L_{1} (W) \left[\frac{1}{3}A_{0} (W) + \frac{2}{3} A_{2} (W) \right] + L_{2} (W) \left[\frac{2}{5} A_{1} (W) + \frac{3}{5} A_{3} (W) \right] + \ldots \mathrm{,} \nonumber \\
 \tilde{A}_{2} (W) &= {L_{0} (W) A_{2} (W) } + L_{1} (W) \left[\frac{2}{5} A_{1} (W) +\frac{3}{5} A_{3} (W) \right] + L_{2} (W) \left[\frac{1}{5} A_{0} (W) + \frac{2}{7} A_{2} (W) + \frac{18}{35} A_{4} (W) \right] + \ldots \mathrm{.}  \nonumber \\
 \vdots \nonumber
\end{align}
%\twocolumngrid
 \\ \noindent
Let us also, word by word, repeat the conclusion given in Ref. \cite{Svarc2018} in which the message essential for this paper is explicitly given (and the important part is for the convenience of the reader emphasized): \\ \\
"The consequence of Eqs.~(\ref{Eq:mixing})~and~(\ref{Eq:mixing-expanded}) is that angular-dependent phase rotations
mix multipoles.
Without fixing the free continuum ambiguity phase $\phi(W,\theta)$,
the partial-wave decomposition $A_\ell (W)$ defined in Eq.~(\ref{Eq:PW}) is non-unique. Partial waves get mixed, and  identification of resonance quantum numbers might be changed.
\emph{\textbf{\small To compare different partial-wave analyses, it is  essential to match  continuum ambiguity phase;
otherwise the mixing of multipoles is yet another, uncontrolled, source of systematic errors.}} Observe that this phase rotation does not create new pole positions, but just reshuffles the existing ones among several partial waves."
\\ \\ \noindent
And this is a starting point of our further analysis. Let us also observe that continuum ambiguity invariance is discussed at the level of amplitudes, and can be applied to any choice of reaction amplitudes whatsoever; hereon we continue our analysis by applying it to one possible choice of reaction amplitudes--helicity amplitudes.
\\ \\ \noindent
 First we in Fig.~\ref{E0+non-rotated} compare the dominant $E0+$ multipole for $\eta$ photoproduction for EtaMAID2018 solution of Mainz EtaMAID model \cite{MAID}, Bonn-Gatchina model \cite{BoGa,BGmultipoles}, Kent State University model \cite{Kent,Manleymultipoles}, J\"{u}lich-Bonn model \cite{Juelich,Doeringmultipoles}  and three solutions\footnote{Solutions I and II are Solutions II and III of Ref. \cite{fixed-t} respectively, and the new yet unpublished solution which is obtained using the same formalism, but multipoles from ref. \cite{MAID} are used for the initial and first constraining solution is denoted as Solution III} from the only SE analysis based on fixed-t constraint \cite{fixed-t}, directly as we get them from original publications, and without paying any attention to the reaction amplitude phase.
The shape is fairly similar, and the sign difference between coupled-channel models on one side, and EtaMAID \& SE solutions on the other can be attributed to the initial assumptions of the model. However, the discrepancies are notable, and important. These figures are well known, and very recently shown in Fig. 4 of Ref. \cite{Kent}\footnote{Some small differences can be seen when one compares  Fig.~1 of this publication and Fig.~4 of ref \cite{Kent}, but this is due to the different version of used solutions.}.
\\ \\ \noindent
As a second step, we perform the synchronization of phases among all models at the level of helicity amplitudes by introducing the following phase rotation:
\be \label{Phaserotation}
\tilde{H}_k^{\rm MD}(W, \theta) &=& H_k^{\rm MD}(W, \theta) \, \cdot \, e^{\di \, \Phi_{H_1}^{\rm BG}(W,\theta) - \, \di \, \Phi_{H_1}^{\rm MD}(W, \theta)} \nonumber  \\
k  & = & 1,\ldots,4
\ee
where $\rm MD$ is the generic notation for Mainz-MAID, Bonn-Gatchina, Kent State University, J\"{u}lich-Bonn, and three fixed-t SE solutions, and $\Phi_{H_1}^{\rm BG}(W,\theta)$ is the phase of helicity amplitude $H_1(W,\theta)$ of Bonn-Gatchina model. In this way we have practically replaced different phases of $H_1(W,\theta)$ amplitude of all  models with only one phase, and this phase is arbitrarily (our convention) chosen to be the one from Bonn-Gatchina model. Then we have multiplied remaining three helicity amplitudes in all models with the same phase factor leaving the set of observables unchanged, and finally compared the rotated multipoles. So, Bonn-Gatchina model results stays untouched as the rotating phase for this model is one, and the overall energy and angle dependent phases of all other models are changed.
\\ \\ \noindent
So, let us summarize the procedure:
\begin{itemize}
     \item We have reconstructed all four helicity amplitudes for all seven models from obtained multipoles.
     \item We have applied the phase rotation defined by Eq.~(\ref{Phaserotation}) to all four helicity amplitudes of all seven models
     \item We have made a partial wave decomposition of rotated sets of amplitudes
     \item We show the final result for rotated $E0+$ multipole in Fig.~2.
\end{itemize}
We stress that we could have taken the phase from any other model, and we could have decided to replace the phase of any out of three remaining helicity amplitudes $H_2(W,\theta)$ - $H_4(W,\theta)$. The conclusion would be the same, but the figure would just have the different phase.
 \begin{figure}[h!] \hspace*{-0.5cm}
\begin{center}
\includegraphics[width=0.8\textwidth]{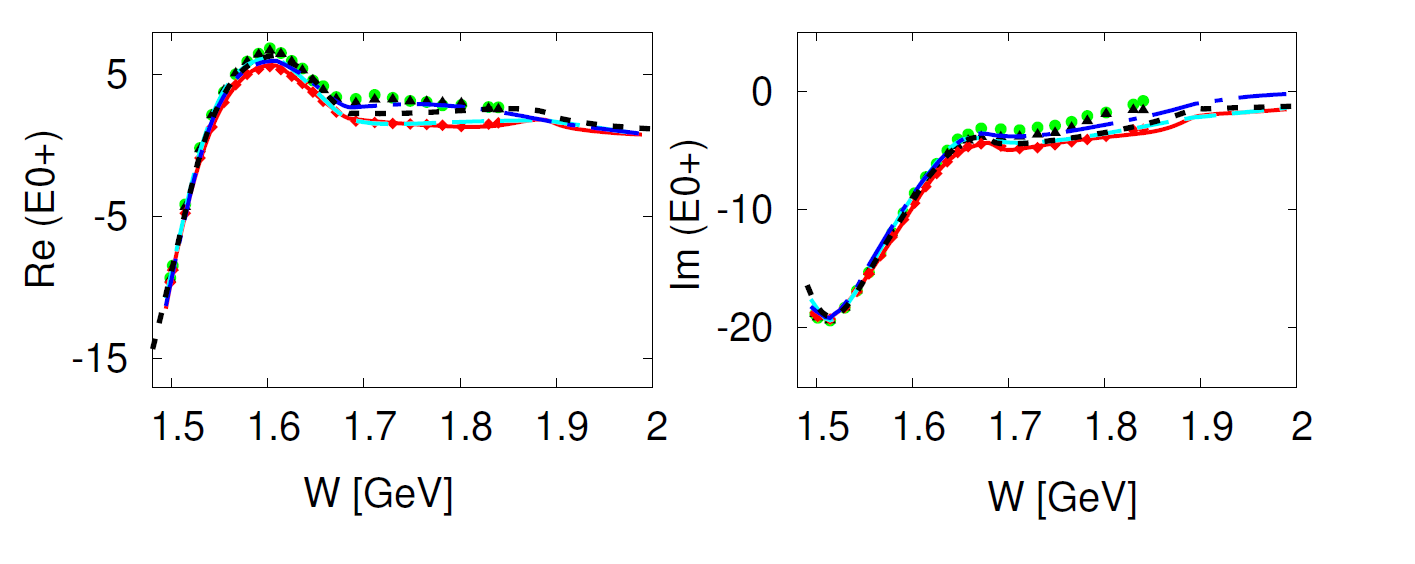}
\caption{\label{E0+rotated}(Color online)  Comparison of $E0+$  $\eta$ photoproduction multipole after the phase rotation defined with Eq.~(\ref{Phaserotation}) for the  Mainz EtaMAID model \cite{MAID},  Bonn-Gatchina calculation \cite{BGmultipoles}, Kent State University calculation \cite{Manleymultipoles},  J\"{u}lich-Bonn model \cite{Doeringmultipoles},  and three solutions from SE fixed-t analysis \cite{fixed-t}. The notation is the same as in Fig.~\ref{E0+non-rotated}.  }
\end{center}
\end{figure}
\\ \\ \noindent
As we claimed, the disagreement among all solutions for  the $E0+$ multipole practically disappeared for energies \mbox{$W_{CM} < $ 1650 MeV}, and it is significantly improved at higher energies.
\\ \\ \indent
Now it is an excellent moment to ask ourselves whether we are at all allowed to touch the phase of reaction amplitudes obtained in coupled-channel calculations. Namely, continuum ambiguity (invariance with respect to the phase rotation) is the consequence of the loss of unitarity. Once the unitarity is restored, continuum ambiguity should disappear.  However, the main aim of coupled-channel models is to restore the unitarity, so the phase ambiguity should be automatically eliminated. Or, a direct consequence should be that all phases of CC ED calculations should be the same, and the phase rotation defined in  Eq.~(\ref{Phaserotation}) should be equal to one.

On the other hand, in Fig.~\ref{E0+rotated}  we  do see that disturbing differences for $E0+$ multipole among all models have disappeared after we applied our phase-rotation synchronization.  This means that the differences seen in  Fig.~\ref{E0+non-rotated} were  the consequence of the mismatch of phases of reaction amplitudes, and that they were not generated either by differences in model assumptions or in data bases chosen. After the phase rotation, the dominant $E0+$ multipole shown in Fig.~\ref{E0+rotated} is up to $\approx 1650$ MeV practically identical for all models and all three SE solutions showing that all options have included sufficient amount of physics to get the unique S-wave result.

\begin{figure}[h!] \hspace*{-0.5cm}
\begin{center}
\includegraphics[width=0.8\textwidth]{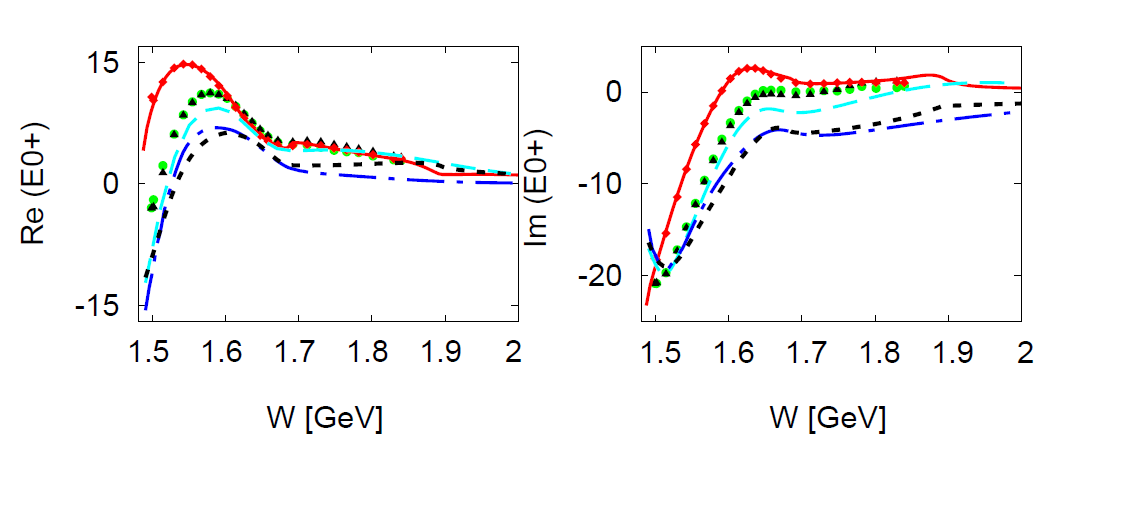}
\caption{\label{E0+xminus1}(Color online)  Comparison of $E0+$  $\eta$ photoproduction multipole of   the Kent State University calculation \cite{Manleymultipoles}, Bonn-Gatchina calculation \cite{BGmultipoles},  J\"{u}lich-Bonn model \cite{Doeringmultipoles}, with   Mainz EtaMAID model \cite{MAID}, and three solutions from SE fixed-t analysis \cite{fixed-t} after the latter four were multiplied with (-1). The notation is the same as in Fig.~\ref{E0+non-rotated}.  }
\end{center}
\end{figure}

This result automatically  confirms that the unitarity in CC calculations is not perfect\footnote{If yes, the phases would be identical, and the phase rotation would be ineffective.}, and that it can only be achieved up to a certain approximation. That is understandable as all channels can never be known, and the treatment of unitarity from calculation to calculation can vary. So, unitarity is only approximately restored, and the phase is only approximately obtained. And this explains why the agreement between ED calculation with a lot of channels (Bonn-Gatchina, J\"{u}lich-Bonn and Kent State University) in Fig.~\ref{E0+non-rotated} is fairly big, and the discrepancy with respect to the calculation where only two channels are included ($\eta$ and $\eta'$) is significant.
\\ \\ \indent
It is important to stress that it might seem that Mainz EtaMAID and all three SE solutions only up to a sign differ from remaining three ED calculations. It is not so. First let us emphasize that the phase of three SE solutions is similar to Mainz EtaMAID solution. The reason for that lies in the mechanism of fixed-t constraining. The fixed-t method is a sophisticated way of fixing the free phase, and it is done by constraining it to the "MAID type"  models. So all three solutions also notably deviate from CC ED calculations, and resemble Mainz EtaMAID type models very much.  That  the  multiplication with minus one is not giving any improvement we show it in Fig.~\ref{E0+xminus1} where the comparison of Mainz EtaMAID ED and three SE solutions multiplied with minus one with remaining ED models is given.

 \begin{figure}[h!] \hspace*{-0.5cm}
\begin{center}
\includegraphics[width=0.85\textwidth]{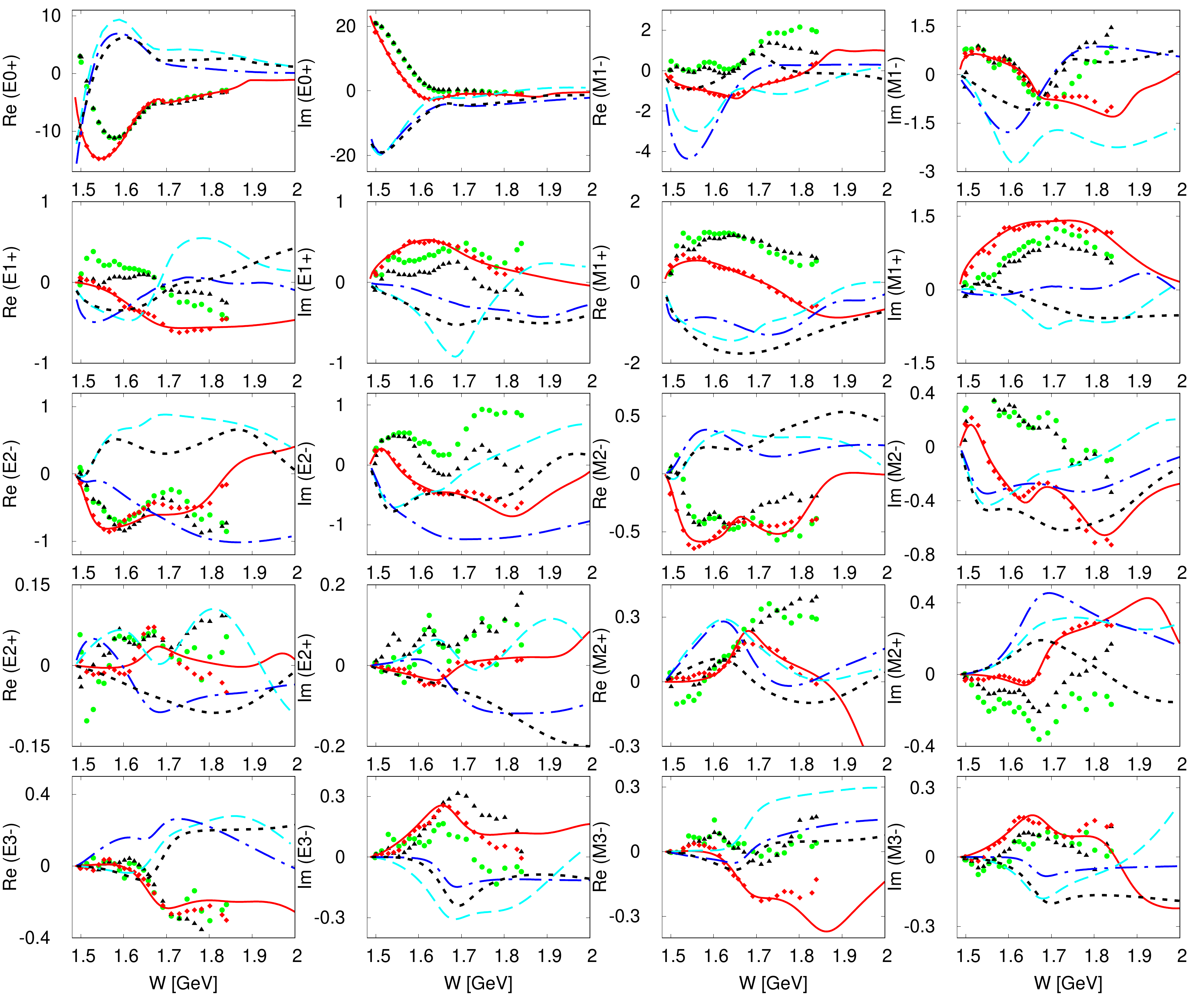} \hspace*{0.5cm}
\caption{\label{non-rotatedmultipoles}(Color online)  Comparison of all $\eta$ photoproduction multipoles  for the Kent State University model \cite{Manleymultipoles}, Bonn-Gatchina model \cite{BGmultipoles},  J\"{u}lich-Bonn model \cite{Doeringmultipoles},  Mainz EtaMAID model \cite{MAID}, and three solutions from SE fixed-t analysis \cite{fixed-t}. The notation is the same as in Fig.~\ref{E0+non-rotated}.  }
\end{center}
\end{figure}

 \begin{figure}[h!] \hspace*{-0.5cm}
\begin{center}
\includegraphics[width=0.85\textwidth]{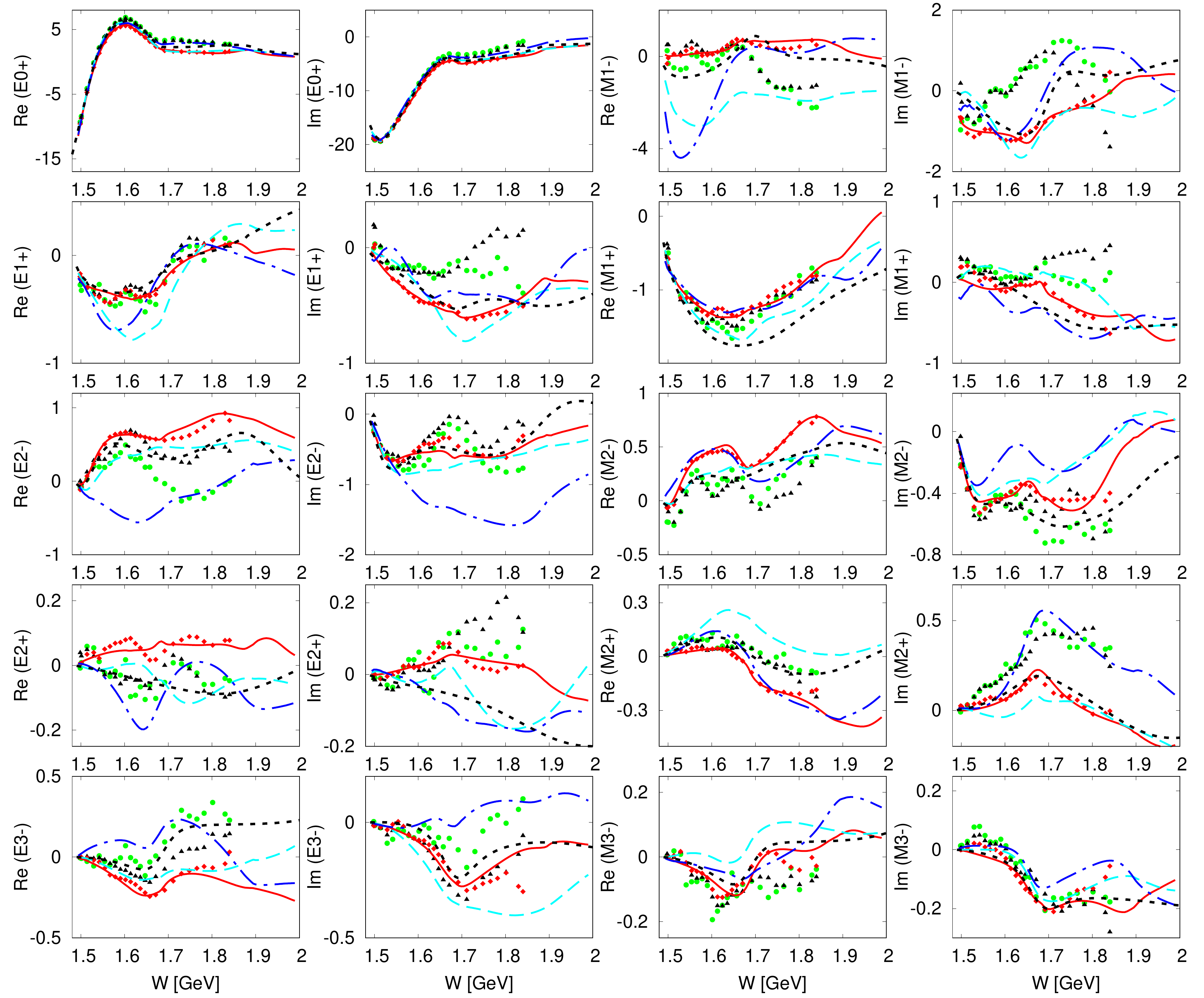}
\caption{\label{rotatedmultipoesl}(Color online)  Comparison of all $\eta$ photoproduction multipoles  after the phase rotation defined with Eq.~(\ref{Phaserotation}) for the Kent State University model \cite{Manleymultipoles}, Bonn-Gatchina model \cite{BGmultipoles},  J\"{u}lich-Bonn model \cite{Doeringmultipoles},  Mainz EtaMAID model \cite{MAID}, and three solutions from SE fixed-t analysis \cite{fixed-t}. The notation is the same as in Fig.~\ref{E0+non-rotated}.  }
\end{center}
\end{figure}

\begin{figure}[h!] \hspace*{-0.5cm}
\begin{center}
\includegraphics[height=5.5cm,width=0.85\textwidth]{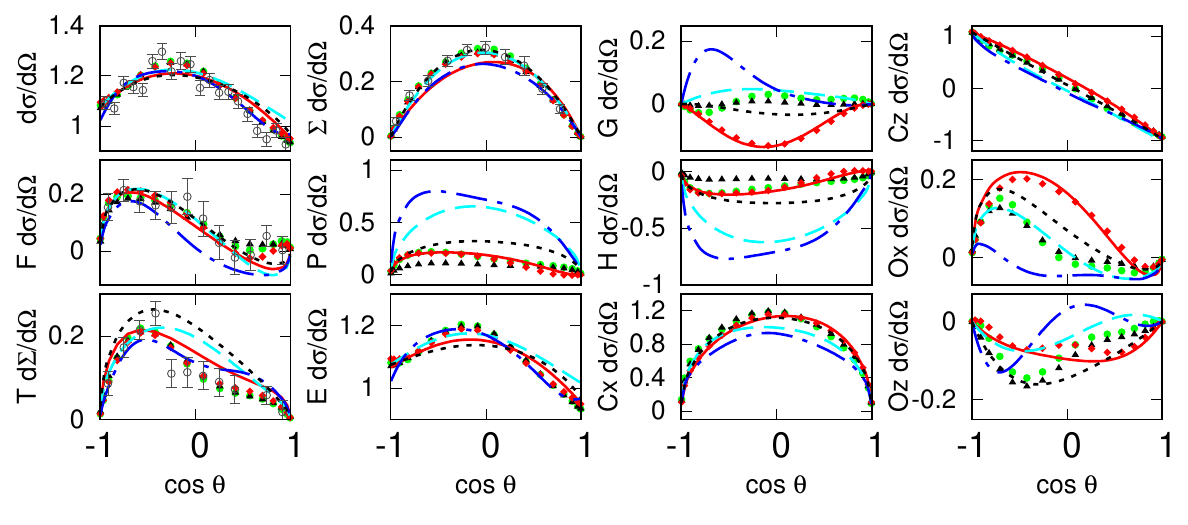} \vspace*{-0.5cm}\\
{\large \textbf{(a)}} \\
\includegraphics[height=5.5cm,width=0.85\textwidth]{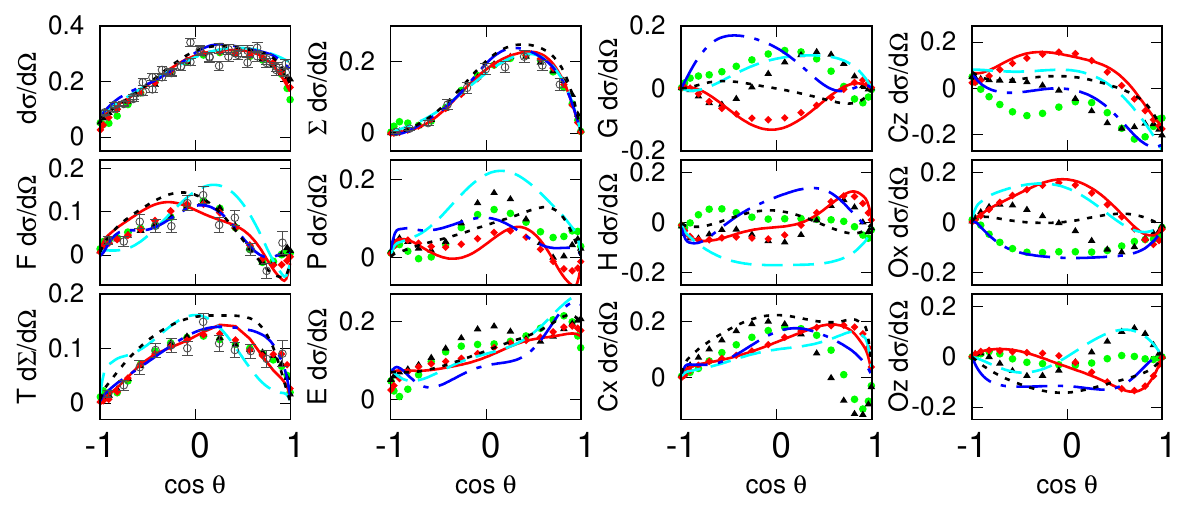} \vspace*{-0.5cm} \\
{\large \textbf{(b)}} \\
\includegraphics[height=5.5cm,width=0.85\textwidth]{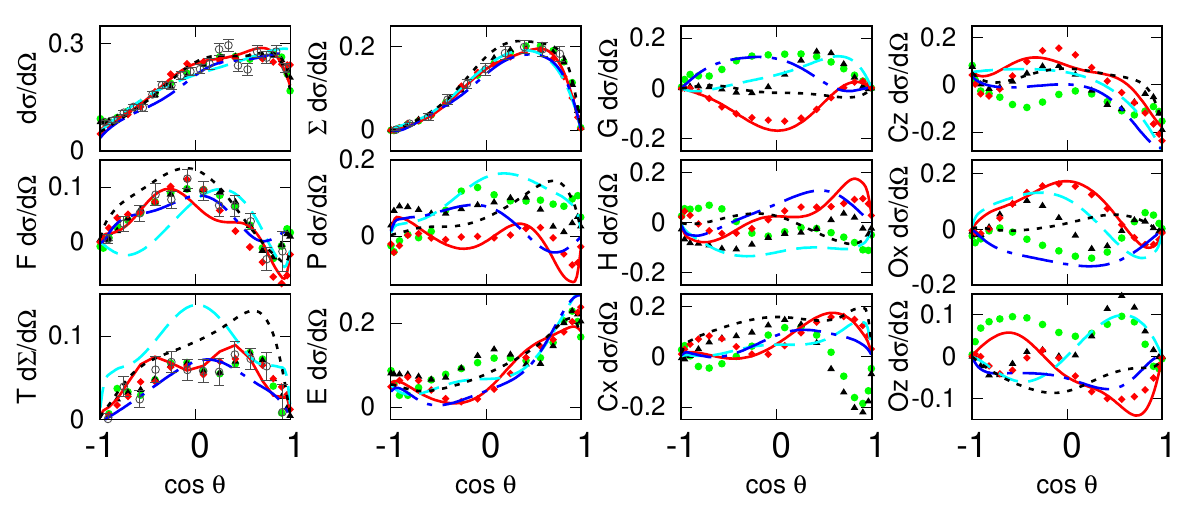} \vspace*{-0.5cm} \\
{\large \textbf{(c)}} \\
\includegraphics[height=5.5cm,width=0.85\textwidth]{FigPred1335.pdf} \vspace*{-0.5cm} \\
{\large \textbf{(d)}}
\caption{\label{Predictions}(Color online) Figs. (a)-(d) show predictions for all seven solutions for 12 measured and unmeasured observables at $W_{CM}=$ 1554, 1602, 1765 and 1840 MeV respectively. Experimental data are shown with grey symbols with error bars, and the notation of all seven model is given in Fig.~\ref{E0+non-rotated}.   }
\end{center}
\end{figure}

\begin{figure}[h!] \hspace*{-0.5cm}
\begin{center}
\includegraphics[width=0.85\textwidth]{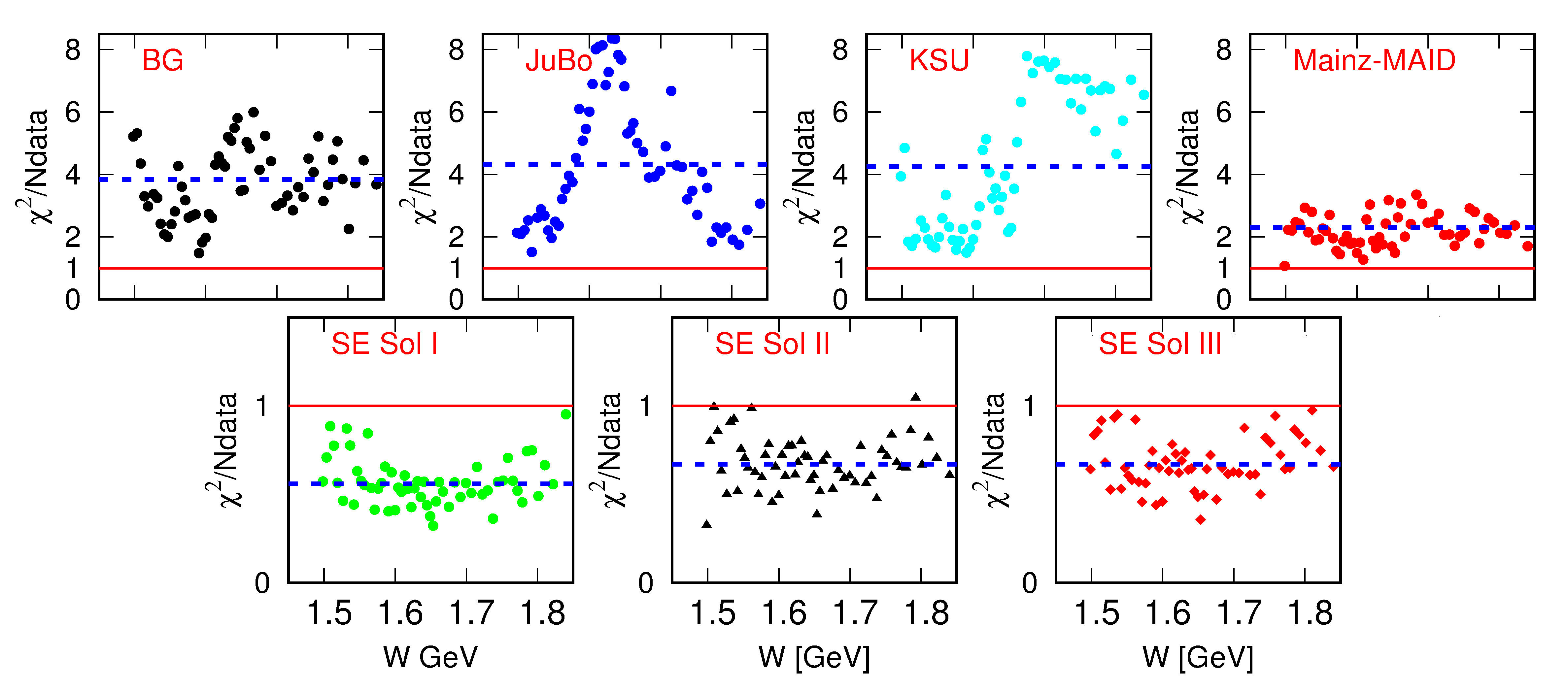} \vspace*{-0.5cm}\\
\caption{\label{Fig6}(Color online) Energy distribution of $\chi^2/N_{data}$ for all seven solutions. Blue dashed line indicates the $\chi^2/N_{data}$ average over all energies. }
\end{center}
\end{figure}

 \noindent
 Synchronizing the phase  on the level of helicity amplitudes, given in Fig.~\ref{E0+rotated}, solves the problem.
 \\ \\ \indent
Comparison of other multipoles is shown in Fig.~3. and in Fig.~4. In Fig.~3 we show the comparison of non-rotated multipoles, exactly as they are given in original publications, and in Fig.~4. we show their comparison after the phase rotation defined by  Eq.~(\ref{Phaserotation}). We see that the grouping of solutions aftre the phase rotation is for some multipoles improved, but no definite consensus can yet be made. So, it seems that we have seven solutions with very similar S-wave, and which are rather different elsewhere. Consequently, the difference should be visible when we show the prediction for all observables from all seven analyzed solutions. The agreement of all solutions with measured observables should be very similar, and for the unmeasured ones it  should be very different. So, in Fig.~\ref{Predictions} we show the prediction for 12 measured and unmeasured observables at four typical energies of \mbox{$W_{CM}=$ 1554, 1602, 1765 and 1840 MeV} for the data base of ref. \cite{fixed-t}. We have used recent A2@MAMI data for unpolarized differential cross section $\sigma_0$, single target polarization asymmetry $T$ and double beam-target polarization with circular polarized photons $F$. In addition we have used the GRAAL data for single beam polarization $\Sigma$. For details, see Table~\ref{tab:expdata}. There are additional data from CLAS~\cite{Dugger:2002ft,Williams:2009yj} and from CBELSA~\cite{Crede:2009zzb}, which we don't use in our analysis. At low energy the cross section data from MAMI have much better statistics. And the higher energies we did not analyze as our fixed-$t$ method becomes more difficult at higher energies.
\begin{table}[htb]
\begin{center}
\caption{\label{tab:expdata} Experimental data from A2@MAMI and
GRAAL used in our PWA.}
\bigskip
\begin{tabular}{|c|c|c|c|c|c|c|}
\hline
 Obs        & $N$ & $E_{lab}$~[MeV] & $N_E$  & $\theta_{cm}$~[$^0$] & $N_\theta$ & Reference    \\
\hline
 $\sigma_0$ & $2400$ & $710 - 1395$ & $120$  & $18 - 162$ & $20$ & A2@MAMI(2010,2017)~\cite{McNicoll:2010qk,Kashevarov2017}  \\
 $\Sigma$   & $ 150$ & $724 - 1472$ & $ 15$  & $40 - 160$ & $10$ & GRAAL(2007)~\cite{Bartalini:2007fg} \\
 $T$        & $ 144$ & $725 - 1350$ & $ 12$  & $24 - 156$ & $12$ & A2@MAMI(2014)~\cite{Annand2014prl} \\
 $F$        & $ 144$ & $725 - 1350$ & $ 12$  & $24 - 156$ & $12$ & A2@MAMI(2014)~\cite{Annand2014prl} \\
\hline
\end{tabular}
\end{center}
\end{table}
\clearpage
We see that the agreement of all seven solutions with measured observables is good, so the reason why non-zero partial waves in all seven solutions differ have to be found in other, non-measured observables which deviate significantly. To quantify this discussion we in Fig.~\ref{Fig6} show the energy distribution of $\chi^2/N_{data}$ for all seven solutions. One has to be very careful not to confuse our numbers with numbers quoted in original publications, because they are produced on the different data base, but the overall trend must be similar.

The results are as to be expected. The best agreement with the data is achieved for the all three SE solutions obtained by the fixed-t analysis.  This is normal as this is single-channel and single-energy analysis which is made continuous by fixing the phase.  Second best agreement is shown by the Mainz EtaMAID analysis which is a two-channel analysis ($\eta$-N and $\eta'$-N channels) with more free parameters per analyzed data then remaining three ED analyses, so this is not a surprise too. The apparently worst result is shown by BG, JuBo and KSU ED analyses, but this was to be expected as they fit much more channels at the same time, and some compromise among channels  has to be made. Due to the coupling with other channels as $\pi$-N, $\sigma$-N, $\rho$-N, $\pi$-$\Delta$, K-$\Lambda$, K-$\Sigma$, $\omega$-N, the BnGa, JuBo and KSU analysis have significantly larger $\chi^2$ values in some energy regions.
What is surprising is the energy dependence of $\chi^2/N_{data}$ in all three ED coupled-channel models, which still awaits fore some explanation.
\\ \\ \indent
As a summary we state that matching angular dependent phases of all solutions on the level of helicity amplitudes brings all $E0+$ multipoles from all seven analyzed PWA into complete agreement. The differences in other partial waves remain. New measurements are needed to fix higher partial waves.

\end{document}